\documentclass[12pt]{article}
\usepackage{bbm}
\def\IZ{\mathbbm Z}

\def\Tr{{\rm Tr}}
\def\and{{\rm and}}
\def\for{{\rm for}}

\def\a{\alpha}
\def\b{\beta}
\def\d{\delta}
\def\g{\gamma}
\def\k{\kappa}
\def\l{\lambda}
\def\s{\sigma}
\def\t{\tau}
\def\om{\omega}

\def\G{\Gamma}
\def\D{\Delta}

\begin{document}
\vspace*{-.6in}
\thispagestyle{empty}
\begin{flushright}
CALT-68-2793\\
\end{flushright}
\baselineskip = 18pt

%\large
%\normalsize

\vspace{1.5in}
{\LARGE
\begin{center}
SOME RECENT PROGRESS IN AdS/CFT\end{center}}
\vspace{.5in}

\begin{center}
John H. Schwarz
\\
\emph{California Institute of Technology\\ Pasadena, CA  91125, USA}
\end{center}
\vspace{.5in}

\begin{center}
\textbf{Abstract}
\end{center}
\begin{quotation}
\noindent Much of modern string theory
research concerns AdS/CFT duality, or more generally, gauge/gravity duality.
The main subjects are

\begin{itemize}

\item Testing and understanding such dualities by exploring how they work for systems with
a lot of supersymmetry

\item Constructing and exploring approximate string theory duals of QCD

\item Applying gauge/gravity duality to other areas of physics such as
condensed matter and nuclear physics

\end{itemize}

\noindent I will briefly discuss the first topic.

\end{quotation}

\vfil
\centerline{Contribution to the proceedings of a symposium}
\centerline{celebrating the 80th birthday of Murray Gell-Mann}

\newpage

\pagenumbering{arabic}

\section{Personal Remarks}

Since this is a conference in honor of Murray Gell-Mann on the occasion of his
80th birthday, it seems appropriate to begin by making some
personal remarks. Murray Gell-Mann has been a very important influence in my
physics career.  This is a good opportunity for me to briefly reminisce about this.

In the fall of 1963, the beginning of my second year of graduate study at UC Berkeley,
a new physics building (Birge Hall) opened. At the time Berkeley (as well as various
other institutions) was trying to enhance its efforts in theoretical particle physics
by hiring Murray Gell-Mann. Murray was already a professor in Caltech, but Berkeley hoped
that he might be enticed into moving. A prime corner office in Birge Hall, with a great view
of the San Francisco Bay and the Berkeley campus, was selected to be Murray's. However,
he hadn't yet decided whether to accept the Berkeley offer, and it was time to occupy the
building. The office could not be left vacant. If another Professor were
assigned to it, it would be awkward to ask him to give it up for Murray. Therefore,
it was decided to assign it to graduate students, who would be much easier to dislodge.
As a result, that was the office that David Gross and I shared for the next three years.
In this way Murray impacted my life before I even met him! Of course, in this period I studied
the eight-fold way (known nowadays as SU(3) flavor symmetry) with great interest.
So he was also influencing me scientifically.

After graduating from Berkeley I spent six years in Princeton, the last three as an
Assistant Professor. In 1972 it was time for me to leave Princeton. The job market at the time
was absolutely terrible. There had been enormous expansion of science faculties for more than
a decade following the launch of Sputnik, but suddenly it came to a screeching halt and there
were almost no jobs. Many good people were driven from the field at that time. My
survival entailed an element of luck. In 1971 Neveu and I discovered a string theory,
which we called the ``dual pion model.'' This, together with Ramond's work on
fermionic strings, led to what is now known as superstring theory. This work
was motivated by the desire to describe hadron physics -- the application to
gravity and unification came later. String theory (called ``dual resonance theory''
in those days) had a couple hundred enthusiastic devotees, but it was still
a relatively small, and somewhat isolated, segment of the
particle theory community. In 1971--72 Murray
was spending a sabbatical year at CERN, which had a strong group of string theorists.
Some of them, such as Brink, Olive, and Scherk, were very interested in supersymmetric
string theory and contributed to its development. Even though Murray was not
working on this himself, he learned of these developments, and decided that
this research could be important. As a consequence, he arranged for me to be
offered a senior research position at Caltech, which I was delighted to accept.

During my first couple years at Caltech, Murray collaborated with Fritzsch and
Minkowski on the development of QCD, and the standard model quickly fell into place.
I followed all this closely, but I continued to work on string theory. Murray
made funds available to me to bring collaborators to Caltech for extended visits. This
facilitated my collaborations with Brink and Scherk, and (much later) with Green.
String theory fell out of favor once it was realized that QCD is the right theory
of hadrons. However, during Scherk's visit in 1974 we realized that string theory
could be used for gravity and unification instead, and this converts several of
its shortcomings into advantages. This change in direction is what convinced me that
it would be worthwhile to continue pursuing the subject.
By then, the community had little interest in
string theory, and it took ten years for this proposal to gain traction. Murray,
however, understood that it could be very important, and so he continued to support
me. I recall him saying that as a committed environmentalist he recognized the
importance of protecting endangered species, and I represented one of them.
I have always felt that he has exceptional judgment in these matters.

In January 1989, nine months before his actual birthday, I organized a celebration
of Murray's 60th birthday at Caltech. There were two days of lectures. The first day
was devoted to physics and the second day to a variety of other subjects in which
Murray was interested. A few years later Murray left Caltech and moved to Santa Fe, where
he was a founding member of the Santa Fe Institute. I have missed seeing him on a regular 
basis, a privilege that I had for about 20 years.

\section{Review of Some Basic Facts}

Let me now turn to gauge/gravity duality.\footnote{The remainder of this manuscript
is very similar to one that I wrote for Shifman's 60th birthday \cite{Schwarz:2009hr}.}
In Maldacena's original paper \cite{Maldacena:1997re}, he proposed
three maximally supersymmetric examples of {AdS/CFT duality}. A basic
indication that the dualities (or equivalences) are plausible
is that the symmetries match. In each case, there is a supergroup,
which describes the isometries of the string theory or M-theory
background geometry. The same supergroup appears as the
superconformal symmetry group of the dual quantum field theory.
Also, the string theory or M-theory solution has $N$ units of flux
threading the sphere factor in the geometry. In fact, the background
configuration corresponds to the near-horizon geometry of $N$
coincident branes, each of which contributes one unit of flux. The
dual conformal field theory, which also depends on the integer $N$,
is the low energy world-volume theory on the branes.

\begin{itemize}
\item {\bf M2-brane Duality:} M-theory on {$AdS_4 \times S^7$}
is dual to a superconformal field theory (SCFT) in three dimensions.
The superconformal symmetry is described by the supergroup is $OSp(8|4)$.

\item {\bf D3-brane Duality:} Type IIB superstring theory on
$AdS_5 \times S^5$ is dual to a SCFT in four dimensions, specifically
${\cal N} =4$ super Yang--Mills (SYM) theory.
The superconformal symmetry in this case is $PSU(2,2|4)$.

\item {\bf M5-brane Duality:} M theory on $AdS_7 \times S^4$
is dual to a SCFT in six dimensions. The superconformal symmetry
in this case is $OSp(6,2|4)$.

\end{itemize}

\subsection{The type IIB / ${\cal N} =4$ SYM example}

This by far the most studied, and best understood, example. The $N$
units of flux ($\int_{S^5} F_5 \approx N$) in the superstring
solution correspond to the gauge group $SU(N)$ in the ${\cal N} =4$
super Yang--Mills theory \cite{Brink:1976bc}. The gauge theory has a
well-known large-$N$ topological ('t Hooft) expansion
\cite{'tHooft:1973jz}. The expansion is in powers of $1/N$ for large
$N$ at fixed $\l$, where the 't Hooft parameter is
\begin{equation}
\l = g_{\rm YM}^2 N.
\end{equation}
This expansion corresponds to the loop expansion of the string theory.
One also identifies
\begin{equation}
 R^2/\a' \approx \sqrt{\l} \quad \and \quad g_{\rm s} \approx \l / N ,
\end{equation}
where $R$ is the radius of the $S^5$ and the $AdS_5$. $g_{\rm s}$ is the
string coupling constant determined by the value of the dilaton field, which
is a massless scalar mode of the string.

\subsection{The type IIA / ABJM example}

There has been significant progress in the last few years in
understanding the M2-brane duality. The suggestion
\cite{Schwarz:2004yj} that the three-dimensional SCFT should be
Chern--Simons gauge theory was implemented for maximal supersymmetry
(${\cal N} =8$) by Bagger and Lambert \cite{Bagger:2007jr} and by
Gustavsson \cite{Gustavsson:2007vu}. However, their construction
only works for the gauge group $SO(4)$, and it does not provide the
desired dual to M-theory on $AdS_4 \times S^7$.

The correct construction was eventually obtained by Aharony,
Bergman, Jafferis, and Maldacena (ABJM) \cite{Aharony:2008ug}. One
key step in their work was to consider a more general problem:
M-theory on $AdS_4 \times S^7/\IZ_k$, with $N$ units of flux. This
gives 3/4 maximal supersymmetry for $k >2$. Thus, the dual gauge
theory is an ${\cal N} =6$ superconformal Chern--Simons theory in
three dimensions. The appropriate gauge group turns out to be
$U(N)_k \times U(N)_{-k}$, where the subscripts are the levels of
the Chern--Simons terms. The ABJM theory also contains bifundamental
scalar and spinor fields. This theory has a topological large-$N$ expansion,
just like the usual ones in four dimensions, for which the 't Hooft
parameter that is held constant in the limit is
\begin{equation}
\l = N/k.
\end{equation}
The only unusual feature is that the 't Hooft parameter is rational.
The extension of the supersymmetry from ${\cal N} =6$ to ${\cal N} =8$
for $k=1,2$ is a nontrivial property of the quantum theory.

The orbifold $S^7/\IZ_k$ can be described as a circle bundle over a
$CP^3$ base. The circle has radius $R/k$, where $R$ is the $S^7$
radius. When $k^5 \gg N$, there is a weakly coupled type IIA
superstring interpretation with string coupling constant
\begin{equation}
g_{\rm s} \approx (N/k^5)^{1/4}.
\end{equation}
One then obtains the correspondences
\begin{equation}
R^2/\a' \approx \sqrt{\l} \quad \and \quad g_{\rm s} \approx \l^{5/4}/N,
\end{equation}
which is very similar to the previous duality. This type IIA duality has
3/4 as much supersymmetry as the type IIB duality, and it is somewhat more
complicated.

\subsection{AdS energies and conformal dimensions}

The geometry of Anti de Sitter space is usually described in Poincar\'e
coordinates, which describes all of the spacetime that is within the light-cone
of a given observer, but does not cover the entire spacetime. For the purpose
of defining energies that correspond to the dimensions of conformal operators,
one needs to use different coordinates, called global coordinates, that cover the
entire spacetime. The metric of $AdS_{p+2}$ in global coordinates is
\begin{equation}
ds^2[AdS_{p+2}] = d\rho^2 - \cosh^2 \rho \, dt^2
+ \sinh^2 \rho \, ds^2[S^p].
\end{equation}
Here, $ds^2[S^p]$ denote the metric of a unit $p$-dimensional
sphere. Actually, AdS/CFT duality requires taking the covering space
of AdS, which means that the global time coordinate $t$ runs from $-\infty$
to $+\infty$.

Witten \cite{Witten:1998qj} and Gubser, Klebanov, Polyakov
\cite{Gubser:1998bc} gave a prescription for relating $n$-point
correlation functions in the gauge theory to corresponding
quantities in the string theory. In the case of two-point functions,
the duality relates the energy $E_A$ of a string state $| A \rangle$
(defined with respect to the global time coordinate $t$),
\begin{equation}
H_{\rm string} | A\rangle = E_A | A \rangle ,
\end{equation}
to the conformal dimensions $\D_A$ of the corresponding
gauge-invariant local operator ${\cal O}_A$ for which
\begin{equation}
\langle {\cal O}_A (x) {\cal O}_B (y)\rangle \approx
\frac{\d_{AB}}{|x-y|^{2\D_A}}.
\end{equation}

Specifically, the duality requires that
\begin{equation}
\D_A (\l, 1/N) = E_A (R^2/\a', g_{\rm s}).
\end{equation}
The 't Hooft expansion of the dimension of ${\cal O}_A$ is
\begin{equation}
\D_A (\l, 1/N) = \D_A^{(0)} + \sum_{g=0}^\infty \frac{1}{N^{2g}}
\sum_{l=1}^\infty \l^l \D_{l,g} .
\end{equation}
$\D_A^{(0)}$ is the classical dimension, and the rest
is called the anomalous dimension.

Almost all studies have focused on the planar approximation, (genus
$g=0$), which is dual to free string theory. This restriction may
make the problem fully tractable, but it is still very challenging.
After all, it would be an extraordinary achievement to solve an
interacting four-dimensional quantum field theory even in the planar
approximation.

\subsection{Approaches to testing the dualities}

Given that it is not possible to completely solve any of these
theories, the question arises how best to test and explore the
workings of AdS/CFT duality. The most obvious things---matching
symmetries and the dimensions of chiral primary operators---have
been done long ago. One wants to dig deeper. One approach is to
match, as much as possible, energies and dimensions of
fields/operators that are not protected by supersymmetry. It should
be noted, however, that a complete test of the duality would also
require matching three-point correlators, since a conformal field
theory is completely characterized by its two-point and three-point
functions. There has been much less progress on this front.

One approach that has been quite successful is the following. First,
identify tractable examples of classical solutions of the string
world-sheet theory. Next, examine the spectrum of small excitations
about these solutions and compute their energies $E_A$. Finally,
identify the corresponding class of operators in the dual gauge
theory and compute their dimensions $\D_A$ in the planar
approximation. Then compare to $E_A$. One subtlety in this analysis
is that this comparison requires an extrapolation from large $\l$,
where the classical world sheet theory is valid, to small $\l$,
where the gauge theory can be studied perturbatively. Thus, one
needs to identify examples in which this is possible. As we will
see, in practice this has conjectural aspects.

A variant of the preceding procedure is to compare equations that
determine $E_A$ and $\D_A$ rather than the solutions. Approaches
based on integrability and algebraic curves try to obtain equations
of ``Bethe type'' on both sides and to match them. This is a very
active area of research, but I will not be able to review it here.
One important issue is that it is much easier to study the
world-sheet theory when the range of $\s$ is infinite (rather than a
circle). In other words, the string itself is infinite, rather than
a loop. In the gauge theory analysis this corresponds to the
thermodynamic limit of the Bethe equations arising from a spin-chain
analysis. There has been progress recently in extending the
integrability techniques to the compact case \cite{Gromov:2009zb}.
However, the story is
quite technical, and I don't think it is completely settled.

\section{Classical String Solutions}

For the reasons outlined above, we want to identify classical string
solutions in the $AdS_5 \times S^5$ background that can be used to
test the duality. The discussion that follows largely follows an
excellent review article by Plefka \cite{Plefka:2005bk}. Other
useful reviews include \cite{Tseytlin:2003ii,Kristjansen:2009}.

The bosonic part of the string world-sheet
action has six cyclic coordinates:
\begin{equation}
(t,\varphi_1 , \varphi_2; \phi_1,\phi_2,\phi_3),
\end{equation}
where the first three coordinates pertain to $AdS_5$ and the second
three to $S^5$. Specifically, we parametrize $S^5$ as follows:
\begin{equation}
ds^2 (S^5) = d\g^2 + \cos^2\g \, d\phi_3^2 + \sin^2\g \, ds^2(S^3),
\end{equation}
where
\begin{equation}
ds^2(S^3) = d\psi^2 +\cos^2\psi \, d \phi_1^2 + \sin^2\psi \, d\phi_2^2.
\end{equation}
Associated to these cyclic coordinates one has conserved charges
\begin{equation}
(E, S_1, S_2; J_1, J_2, J_3).
\end{equation}
$E$ is the energy and the other five charges are angular momenta.

One much-studied class of string solutions involves a line up the
center of $AdS_5$, described by $\rho = 0$ and $t =\k \t$, where
$\k$ is a constant and $\t$ is the world-sheet time coordinate.
These configurations have $S_1 =S_2 =0$.

\subsection{Point-particle solutions}

The simplest solution is a point particle (collapsed string)
encircling the sphere. In addition to $\rho = 0$ and $t =\k \t$,
this is described by
\begin{equation}
\g =\pi/2, \quad  \phi_1 =\k\t,  \quad \psi=0.
\end{equation}
This has $J_2 =J_3 =0$.

The quantum excitations of this solution have energies that can be
expanded in powers of $1/J$ for large $J=J_1$, where
\begin{equation}
\k = J/\sqrt{\l}
\end{equation}
is held fixed. This is equivalent to the BMN analysis of strings in a
plane-wave background \cite{Berenstein:2002jq}. One obtains
\begin{equation}
E-J \approx E_2(\k) + \frac{1}{J} E_4(\k) + \ldots
\end{equation}

The exact BMN result is
\begin{equation}
E_2 = \sum_{n= -\infty}^{\infty} \sqrt{n^2 + \k^2} N_n,
\end{equation}
where $N_n = \sum_{i=1}^8 \a_n^{i\dagger} \a_n^i + {\rm fermions}$
is expressed in terms of ordinary oscillators
\begin{equation}
[\a_m^i,\a_n^{j\dagger}] = \d^{ij} \d_{mn}.
\end{equation}
The level matching condition is $\sum n N_n =0$.

The BMN paper proposed a scaling rule, known as BMN scaling, which
predicts agreement with the anomalous dimensions of operators in the
dual gauge theory, even though one calculation is valid for large
$\l$ and the other for small $\l$. In other words, their scaling
hypothesis, if valid, would justify the extrapolation from small
$\l$ to large $\l$. In fact, it turns out that $E_2$ agrees
perfectly, but agreement for $E_4$ breaks down at three loops
\cite{Callan:2003xr}. This is not a problem for AdS/CFT duality,
only for the BMN scaling conjecture.\footnote{Perhaps it would be
more fair to say that the BMN scaling conjecture was made for the
plane-wave limit only, which corresponds to $E_2$; what fails is an
attempt to generalize the scaling conjecture beyond that.}

\subsection{Spinning string solutions}

A class of interesting generalizations of the preceding solution
describes circular or folded strings that are extended on the $S^3
\subset S^5$. These have $t = \k\t$, $\rho = 0$, and $\g = \pi/2$,
as before. But now one takes
\begin{equation}
\phi_1 = \om_1 \t, \quad \phi_2 =
\om_2 \t, \quad \psi = \psi(\s).
\end{equation}
For these choices, the string equation of motion gives
\begin{equation}
\psi'' + \om_{21}^2  \sin\psi \cos\psi =0,
\end{equation}
where $\om_{21}^2 = \om_2^2 - \om_1^2$. This is the well-known
pendulum equation.

This equation has a first integral
\begin{equation}
\psi' = \om_{21} \sqrt{q-\sin^2\psi}, \quad q= (\k^2
-\om_1^2)/\om_{21}^2.
\end{equation}
The solution for  $q<1$, which involves the elliptic integrals
$E(q)$ and $K(q)$, describes a folded string. It corresponds to a
pendulum that oscillates back and forth. The solution for $q>1$,
which involves the elliptic integrals $E(q^{-1})$ and $K(q^{-1})$,
describes a circular string. It corresponds to a pendulum that goes
round and round. In the classical limit, the energy has the form
\begin{equation}
E = \sqrt{\l} F( J_1/\sqrt{\l}, J_2/\sqrt{\l}).
\end{equation}

\subsection{Dual gauge theory analysis}

This string theory result can be extrapolated to small $\l$ and
compared to the dual gauge theory.  The operators that carry $J_1,
J_2$ charges have the form
\begin{equation}
{\cal O}_\a^{J_1, J_2} = \Tr\left( Z^{J_1} W^{J_2}\right) + \ldots
\end{equation}
where $Z$ and $W$ are complex scalar fields in the adjoint of $SU(N)$.
The additional terms denoted by dots involve different orderings of the
$Z$s and $W$s. Such a trace can be viewed as a ring configuration of an
$S=1/2$ quantum spin chain, where $W$ corresponds to spin up and $Z$
corresponds to spin down.

The conformal dimensions of operators ${\cal O}_\a^{J_1, J_2}(x)$
with these charges are eigenvalues of the dilatation operator
\begin{equation}
{\cal D} {\cal O}_\a^{J_1, J_2}(x) = \sum _\b D_{\a\b} {\cal
O}_\b^{J_1, J_2}(x).
\end{equation}
In the planar one-loop approximation the equations are precisely
those of a ferromagnetic Heisenberg spin chain, which is a
well-known integrable system, whose Hamiltonian is proportional to
\begin{equation}
{\cal H} = \sum _{i=1}^J \left(\frac{1}{4} -
\vec\s_i\cdot\vec\s_{i+1}\right).
\end{equation}
This can be solved using Bethe ansatz techniques, thereby obtaining
conformal dimensions that can be compared (successfully) with
energies of the corresponding string solutions. Higher-order terms,
which correspond to more complicated spin-chain Hamiltonians, have
also been studied.

\subsection{Strings spinning in AdS}

Another interesting class of classical string solutions are ones in
which the string position is extended in the AdS space and a
point moving on the sphere. The first example of this type is the
straight folded string rotating in $AdS_3 \subset AdS_5$
\cite{Gubser:2002tv}. One finds that for large $S$
\begin{equation}
E= 2\G(\l) \log S + O(S^0),
\end{equation}
where
\begin{equation}
\G(\l) = \frac{\sqrt{\l}}{2\pi} + O(\l^0) \quad \for \quad \l \gg 1.
\end{equation}

The dual gauge theory operators are
\begin{equation}
\Tr (D_+^{s_1} Z \, D_+^{s_2} Z) \quad s_1 + s_2 =S.
\end{equation}
Their anomalous dimensions take the same form as the energy with
\begin{equation}
\G(\l) = \frac{\l}{4\pi^2} + O(\l^2)  \quad \for \quad \l \ll 1.
\end{equation}
In order to compare these, one needs a procedure to extrapolate
between small and large $\l$. In fact, an exact formula for the {\em
cusp anomalous dimension} $\G(\l)$ has been deduced using the
assumption of exact integrability \cite{Beisert:2006ez}. It passes
all tests and is likely to be correct.

The generalization of this duality to twist $J$ operators, which
have the form
\begin{equation}
\Tr (D_+^{s_1} Z \, D_+^{s_2} Z \ldots D_+^{s_J} Z) \quad {\rm
where} \quad\sum s_l=S,
\end{equation}
has been explored by Dorey and Losi \cite{Dorey:2008vp}. They
computed the corresponding conformal dimensions using an $SL(2)$
spin chain model. For large $S$ the correspond classical string
solutions are {\em spiky strings} with $J$ cusps. The duality
predictions are verified to the extent that they have been explored.

\section{Conclusion}
There has been a lot of progress in testing AdS/CFT in various
special cases for maximally supersymmetric theories. Much of this
progress has exploited the integrability of the string world-sheet
theory on the one hand and the integrability of various spin-chain
models that arise in studies of the dual gauge theory in the planar
approximation on the other hand. More recently, there
has also been very interesting work exploring analogous
constructions for the M2-brane duality following the discovery of
the ABJM theory. Much less is known about the M5-brane theory,
though there has been significant progress when two of the
dimensions wrap a Riemann surface \cite{Gaiotto:2009we,Gaiotto:2009gz}.
The superconformal theory on flat M5-branes is strongly coupled, and it
does not appear to have a Lagrangian description. Moreover, it seems
to involve tensionless strings, which is a poorly understood subject.

It has been fun traveling half way around the world to the amazing
country of Singapore in order to celebrate my long-time friend and colleague
Murray Gell-Mann. I am looking forward to his 100th birthday celebration.

\section*{Acknowledgments}

This work was supported in part by the U.S. Dept. of
Energy under Grant No. DE-FG03-92-ER40701.


\newpage

\begin{thebibliography}{99}

%\cite{Schwarz:2009hr}
\bibitem{Schwarz:2009hr}
  J.~H.~Schwarz,
  ``Recent Progress in AdS/CFT,''
  Int.\ J.\ Mod.\ Phys.\  {\bf A25}, 310-318 (2010).
  [arXiv:0907.4972 [hep-th]].

%\cite{Maldacena:1997re}
\bibitem{Maldacena:1997re}
  J.~M.~Maldacena,
  ``The Large N Limit of Superconformal Field Theories and Supergravity,''
  Adv.\ Theor.\ Math.\ Phys.\  {\bf 2}, 231 (1998)
  [Int.\ J.\ Theor.\ Phys.\  {\bf 38}, 1113 (1999)]
  [arXiv:hep-th/9711200].
  %%CITATION = IJTPB,38,1113;%%

%\cite{Brink:1976bc}
\bibitem{Brink:1976bc}
  L.~Brink, J.~H.~Schwarz and J.~Scherk,
  ``Supersymmetric Yang--Mills Theories,''
  Nucl.\ Phys.\  B {\bf 121}, 77 (1977).
  %%CITATION = NUPHA,B121,77;%%

%\cite{'tHooft:1973jz}
\bibitem{'tHooft:1973jz}
  G.~'t Hooft,
  ``A Planar Diagram Theory for Strong Interactions,''
  Nucl.\ Phys.\  B {\bf 72}, 461 (1974).
  %%CITATION = NUPHA,B72,461;%%

%\cite{Schwarz:2004yj}
\bibitem{Schwarz:2004yj}
  J.~H.~Schwarz,
  ``Superconformal Chern--Simons Theories,''
  JHEP {\bf 0411}, 078 (2004)
  [arXiv:hep-th/0411077].
  %%CITATION = JHEPA,0411,078;%%

%\cite{Bagger:2007jr}
\bibitem{Bagger:2007jr}
  J.~Bagger and N.~Lambert,
  ``Gauge Symmetry and Supersymmetry of Multiple M2-Branes,''
  Phys.\ Rev.\  D {\bf 77}, 065008 (2008)
  [arXiv:0711.0955 [hep-th]].
  %%CITATION = PHRVA,D77,065008;%%

%\cite{Gustavsson:2007vu}
\bibitem{Gustavsson:2007vu}
  A.~Gustavsson,
  ``Algebraic Structures on Parallel M2-Branes,''
  Nucl.\ Phys.\  B {\bf 811}, 66 (2009)
  [arXiv:0709.1260 [hep-th]].
  %%CITATION = NUPHA,B811,66;%%

%\cite{Aharony:2008ug}
\bibitem{Aharony:2008ug}
  O.~Aharony, O.~Bergman, D.~L.~Jafferis and J.~M.~Maldacena,
  ``$N=6$ Superconformal Chern--Simons-Matter Theories, M2-Branes and Their
  Gravity Duals,''
  JHEP {\bf 0810}, 091 (2008)
  [arXiv:0806.1218 [hep-th]].
  %%CITATION = JHEPA,0810,091;%%

%\cite{Witten:1998qj}
\bibitem{Witten:1998qj}
  E.~Witten,
  ``Anti de Sitter Space and Holography,''
  Adv.\ Theor.\ Math.\ Phys.\  {\bf 2}, 253 (1998)
  [arXiv:hep-th/9802150].
  %%CITATION = 00203,2,253;%%

%\cite{Gubser:1998bc}
\bibitem{Gubser:1998bc}
  S.~S.~Gubser, I.~R.~Klebanov and A.~M.~Polyakov,
  ``Gauge Theory Correlators from Non-critical String Theory,''
  Phys.\ Lett.\  B {\bf 428}, 105 (1998)
  [arXiv:hep-th/9802109].
  %%CITATION = PHLTA,B428,105;%%

%\cite{Gromov:2009zb}
\bibitem{Gromov:2009zb}
  N.~Gromov, V.~Kazakov and P.~Vieira,
  ``Exact AdS/CFT Spectrum: Konishi Dimension at any Coupling,''
  arXiv:0906.4240 [hep-th].
  %%CITATION = ARXIV:0906.4240;%%

%\cite{Plefka:2005bk}
\bibitem{Plefka:2005bk}
  J.~Plefka,
  ``Spinning Strings and Integrable Spin Chains in the AdS/CFT
  Correspondence,''
  Living Rev.\ Rel.\  {\bf 8}, 9 (2005)
  [arXiv:hep-th/0507136].

%\cite{Tseytlin:2003ii}
\bibitem{Tseytlin:2003ii}
  A.~A.~Tseytlin,
  ``Spinning Strings and AdS/CFT Duality,''
  arXiv:hep-th/0311139.
  %%CITATION = HEP-TH/0311139;%%

%\cite{Kristjansen:2009}
\bibitem{Kristjansen:2009}
C. Kristjansen, M. Staudacher, and A. Tseytlin, eds., {\it Special
Issue on Integrability and the AdS/CFT Correspondence}, J. Phys. A
{\bf 42}, 250301 (2009).

%\cite{Berenstein:2002jq}
\bibitem{Berenstein:2002jq}
  D.~E.~Berenstein, J.~M.~Maldacena and H.~S.~Nastase,
  ``Strings in Flat Space and pp Waves from $N = 4$ Super Yang Mills,''
  JHEP {\bf 0204}, 013 (2002)
  [arXiv:hep-th/0202021].
  %%CITATION = JHEPA,0204,013;%%

%\cite{Callan:2003xr}
\bibitem{Callan:2003xr}
  C.~G.~Callan, H.~K.~Lee, T.~McLoughlin, J.~H.~Schwarz, I.~Swanson and X.~Wu,
  ``Quantizing String Theory in AdS(5) x S**5: Beyond the pp-Wave,''
  Nucl.\ Phys.\  B {\bf 673}, 3 (2003)
  [arXiv:hep-th/0307032].
  %%CITATION = NUPHA,B673,3;%%

%\cite{Gubser:2002tv}
\bibitem{Gubser:2002tv}
  S.~S.~Gubser, I.~R.~Klebanov and A.~M.~Polyakov,
  ``A Semi-Classical Limit of the Gauge/String Correspondence,''
  Nucl.\ Phys.\  B {\bf 636}, 99 (2002)
  [arXiv:hep-th/0204051].
  %%CITATION = NUPHA,B636,99;%%

%\cite{Beisert:2006ez}
\bibitem{Beisert:2006ez}
  N.~Beisert, B.~Eden and M.~Staudacher,
  ``Transcendentality and Crossing,''
  J.\ Stat.\ Mech.\  {\bf 0701}, P021 (2007)
  [arXiv:hep-th/0610251].
  %%CITATION = JSTAT,0701,P021;%%

%\cite{Dorey:2008vp}
\bibitem{Dorey:2008vp}
  N.~Dorey and M.~Losi,
  ``Spiky Strings and Spin Chains,''
  arXiv:0812.1704 [hep-th].
  %%CITATION = ARXIV:0812.1704;%%

%\cite{Gaiotto:2009we}
\bibitem{Gaiotto:2009we}
  D.~Gaiotto,
  ``$N=2$ Dualities,''
  arXiv:0904.2715 [hep-th].
  %%CITATION = ARXIV:0904.2715;%%

%\cite{Gaiotto:2009gz}
\bibitem{Gaiotto:2009gz}
  D.~Gaiotto and J.~M.~Maldacena,
  ``The Gravity Duals of $N=2$ Superconformal Field Theories,''
  arXiv:0904.4466 [hep-th].
  %%CITATION = ARXIV:0904.4466;%%


\end{thebibliography}
\end{document}